\documentclass[a4paper,20pt,oneside,Palatino]{article}
	\usepackage{latexsym}
	\usepackage[dvips]{graphics}
	\usepackage[dvips]{color}
	\usepackage{graphicx}
	\usepackage{slashed}
	\usepackage{amssymb}
	\usepackage{amsmath}
	\usepackage{cite}
	\usepackage{simplemargins}
	\setleftmargin{1in}
        \setrightmargin{1in}
        \settopmargin{1in}
        \setbottommargin{0.7in}

\begin{document}
	\begin{flushleft}
	{\Large\textsf{Axion luminosity of Active Galactic Nuclei}}
	\end{flushleft}

	\begin{flushleft}		
	\bigskip
	Pankaj Jain$^{\dag}$\\ 
	
	Physics Department\\
	I.I.T. Kanpur\\
	Kanpur-208016, India \\

\bigskip
	Subhayan Mandal$^{\ddag}$\\

	Fundamental Interactions in
	   Physics and Astrophysics\\
University of Li\`ege \\
all\'ee du 6 ao\^ut, 17
B-4000 Li\`ege 1
Belgium \\
	email:pkjain@iitk.ac.in$^{\dag}$, smandal@ulg.ac.be$^{\ddag}$
	\end{flushleft}
	
	\bigskip
	\noindent
	\begin{flushright}
	\begin{minipage}[c]{0.85\textwidth}
	{\bf Abstract:} We compute the flux of axions from Active Galactic
Nuclei (AGN). Axions can be produced in the accretion disk by the Compton, 
Bremsstrahlung \& Primakoff processes. We find that the axion luminosity
due to these processes is negligible in comparison to the photon luminosity
from AGNs. We also compute the luminosity of a hypothetical pseudoscalar, 
with very small mass, from the AGN atmosphere due to the phenomenon of
pseudoscalar-photon mixing in background magnetic field. In this case we find 
that for some parameter ranges, the pseudoscalar flux can exceed that of 
photons. We comment on the implications of this result on the observed large
scale alignment of optical polarizations from AGNs.
	\end{minipage}
	\end{flushright}

\section{Introduction}
\par\noindent The cores of Active Galactic Nuclei (AGN), identified 
as quasars, emit a huge amount of power at visible and ultraviolet frequencies
\cite{bmp}.
It obtains its power by the gravitational potential energy of a massive
black hole residing at its center \cite{kori}. The radiation is emitted by the accretion
disk surrounding the black hole. 
In this paper we compute the luminosity of the invisible axions 
\cite{PQ1,PQ2,Weinberg,Wilczek,McKay,Kim1,Dine,Zhitnitsky,McKay2,Kim} 
from AGNs. 
We also consider a hypothetical light 
pseudoscalar whose couplings and mass
are not related to one another.  
Our motivation
for this study is two folds. The pseudoscalar flux from AGNs may be used 
to impose limits on its mass and couplings. If the pseudoscalar flux is
sufficiently large then it might also provide an explanation for the
observed large scale alignment of visible polarizations from quasars. 
Large scale alignment, on distance scales of a Gpc, has been observed 
in many regions of the sky \cite{huts98,huts01,huts02,JNS04}. 
A statistically significant signal of alignment
with the local supercluster has also been observed 
\cite{huts98,huts01,huts02,JNS04,JP04}. 
This effect may be explained in terms of the conversion of photons to 
pseudoscalars in the local supercluster magnetic field. However this 
explanation is not consistent with data. The problem arises due to the 
observed difference in the distribution of polarizations among the Radio 
Quiet (RQ) and optically selected (O) quasars and the Broad Absorption 
Line (BAL) quasars. The polarization distribution 
of the RQ and O quasars peaks at very low values. The magnitude
of the mixing required to explain the alignment effect is sufficiently large 
so as to completely wash out this difference. 
In Ref. \cite{JPS02} it was suggested that if the pseudoscalar
flux from quasars is sufficiently large at visible 
frequencies, than conversion in the supercluster magnetic field may 
consistently explain the alignment with supercluster. 
In this case the alignment is explained in terms of the conversion of 
pseudoscalars to photons. 

We first study the emission of pseudoscalars from the accretion disk via  
the Compton, Bremsstrahlung and the Primakoff channels. In this calculation
we assume the pseudoscalar to be the standard axion. Besides emission from
the accretion disk, pseudoscalars may also be produced in the AGN atmospheres
due to the conversion of photons to pseudoscalar in the background magnetic field.
The probability for this conversion is negligible for the standard axion but
can be large if the pseudoscalar mass is very small.

\section{Axion Luminosity from the Accretion Disk}
The emission rates of the axion via the Compton and Bremsstrahlung \& Primakoff channels are given, respectively, by the formulas 
\cite{kolb,fukugita,kmw,Dicus1},
\begin{equation}
{\dot{\epsilon_a}(C) = \frac{40N_A\alpha g_{aee}^2 \zeta (6)T^6}{\mu _e \pi^2 m_e^4}}
\end{equation}
	\begin{equation}
	{\dot{\epsilon_a}(B) = \frac{64n_en_Z g_{aee}^2 Z^2 \alpha ^2T^{5/2}}{15\rho  (2\pi)^{3/2} m_e^{7/2}}}
	\end{equation}
	\begin{equation}
	{\dot{\epsilon_a}(P) = \frac{2n_Z Z^2 \alpha g_{a \gamma \gamma}^2 T^4
\left(6\zeta (4)\left[\ln(2)-0.5-\ln(\frac{\omega_p}{T})\right]+7.74\right)}{\rho \pi}}\, .	 
	\end{equation}
Here ${N_A}$ is Avogadro's number, $\alpha$ is the fine structure constant, 
${\zeta(n)}$ is the Riemann 
zeta function, ${T}$ is the temperature in the accretion disk, 
${\mu_e}$ is the mean molecular weight of electron, ${m_e}$ is 
the electron mass, ${\rho}$ is the density, $n_{e}$ $(n_Z)$ 
is the number density of electrons (nucleons), ${Z}$ is atomic number and 
$\omega_p$ is the plasma frequency.
The axion-electron coupling $g_{aee}$ and the axion-photon coupling,
$g_{a\gamma\gamma}$, is related to the Peccei-Quinn (PQ) spontaneous symmetry
breaking scale, $f_{PQ}$, by the standard formulas \cite{kolb}. We assume
the DFSZ \cite{Dine,Zhitnitsky} axion for which  
$f_{PQ}$ is constrained by observations to be
greater than $ 10^{8}$ GeV \cite{PDG,ggrflt}. For this limiting value the couplings
are found to be, $g_{aee} = 5.0 \times 10^{-12}$ and
$g_{a \gamma \gamma} = 
8.4 \times 10^{-12}\ {\rm GeV}^{-1}$. In these estimates we have set the
color anomaly factor to be unity. Although the values of these parameters 
are model dependent, here we use these values for our estimates.

We next calculate the luminosity of axions from the accretion disk by 
integrating the emission rates over the disk mass.
We assume the thin disk model of the accretion disk. Let $\rho$ denote the 
density of the disk. It can be replaced by $ {\frac{\Sigma}{H}}$, 
where ${\Sigma}$ is the surface density of the disk and 
$H = {R^{3/2}C_s\over\sqrt{GM}}$. Here $C_s=10^6$ cm/s is the speed of 
sound in the accretion disk medium, $G$ is the Gravitation constant, 
and $M$ is the mass of the central black hole, roughly equal to 
$10^{41} {\rm gm}$ \cite{king}.

\subsection{Luminosity due to Compton Scattering}
We first compute the luminosity due to Compton scattering. From the formula
for the emission rate we get,
$\dot{\epsilon_a}(C) = 1.268\times 10^{-11}T^{6}~{\rm GeV}$. 
The luminosity is given by,
\begin{equation}
{L_{comp} = \int\dot{\epsilon_a}(C)\,d M = 1.72 \times 10^{34} \int \int \int \rho T^6 R\,d R \,d \phi\,d z  ~~{\rm erg-s^{-1}}}
\end{equation}
The `z' integration is straightforward, 
$\int dz=H$, where `z' is the scale height of the disk.
	We obtain 
	\begin{equation}
	{L_{comp} = 1.72 \times 10^{34} \times 2\pi\int_{R_{\ast}}^{10^{3}R_{\ast}} \Sigma T^6 R \,d R ~~{\rm erg-s^{-1}}} 
	\end{equation}
	where ${\Sigma}$ and $T$ are given by \cite{king}, 
	\begin{equation}
	{\Sigma = 3.57 \times 10^{33} \left[ \frac{1}{R^{3/2}} - \frac{\sqrt R_{\ast}}{R^2} \right]~~{\rm erg-cm^{-2}}}\,,
	\label{eqn:surfden}
	\end{equation}
	\begin{equation}
{T = 1.1 \times 10^{16} \left[ \frac{1}{R^3} - \frac{\sqrt R_{\ast}}{R^{7/2}} \right]^{1/4}~~{\rm K}}\,,
\label{eqn:temp}
\end{equation}
respectively.
Using these values, we get, $ L_{comp} = 9.7 \times 10^{29} ~~{\rm erg-s^{-1}}$.
	
\subsection{Bremsstrahlung}
The axion emission rate for the bremsstrahlung process is found to be 
\begin{equation}
{\dot{\epsilon_a}(B) = 1.461\times10^{-16} \rho T^{5/2}~~{\rm GeV}}
\end{equation}
where both $\rho$ and T are in GeV units. Using Eqs. (\ref{eqn:surfden}) and (\ref{eqn:temp}) we find, 
\begin{equation}       
{\dot{\epsilon_a}(B) = 1.28\times10^{-22} \frac{\Sigma}{R^{3/2}} T^{5/2}~~{\rm GeV}}\,.
\end{equation} 
The luminosity due to this process is found to be 
\begin{equation}       
{L_{brem}} = \int\dot{\epsilon_a}(B)\,d M = 5.7 \times 10^{36} ~~{\rm erg-s^{-1}}\,.
\end{equation}

\subsection{Primakoff}
In this case, the emission rate is given by,
\begin{equation}
{\dot{\epsilon_a}(P) = 3.4567\times 10^{-25} \left[8.9943 T^4 - 6.4939T^4 \ln \left({\omega_p \over T}\right)\right] ~~{\rm GeV}}
	\end{equation}
where the plasma frequency,
\begin{equation}
{\omega_p = 2.452\times 10^{13} {1 \over R^{3/4}} \left [{1 \over R^{3/2}} - {\sqrt{R_{\ast}}\over R^2}\right]^{1/2} ~~{\rm GeV}}\,.
\end{equation}
	Therefore, we find that, 
	\begin{equation}
	{L_{Prim} = 2.84 \times 10^{32} + 7.1 \times 10^{31} ~~~{ \rm erg - s^{-1}}}
	\end{equation}

We find that the axion emission rate is relatively small due to all the three processes. It is negligible compared to the AGN's photon luminosity. The dominant
contribution is obtained by the bremsstrahlung process. 

The calculation in this section depends only on the coupling of the axion
to fermions and photons and does not depend on its mass. Hence the 
result is also valid for a hypothetical pseudoscalar with similar couplings
but whose mass may not be related to the PQ symmetry breaking scale as long
as the mass is much smaller than the accretion disk temperature. 

We note that in our study we have made several approximations. For 
example, we have used the thin disk approximation in our study, which may not be true in reality. Actually, our usage of the thin disk model is forced since no other viable stable model is available. Furthermore we have not investigated 
the pseudoscalar emission rate from the interiors of the AGN's or from other 
parts such as jets, etc.  

\section{Conversion In AGN Surroundings}
In section 2 we have found that the total axion luminosity of the accretion
disk is negligible compared to the total photon luminosity. Here, we 
determine the contribution to the pseudoscalar luminosity due to photon
to pseudoscalar conversion in the AGN surroundings due to the background 
magnetic field. By AGN surroundings, we mean the atmosphere outside the 
accretion disk. This includes the dust tori, broad line region and
the narrow line region. In order to perform this calculation, we require parameters such as the plasma density, magnetic field etc. in this region, which are unknown. Hence, we instead use the parameters corresponding to the radio lobes. 
We shall take the representative values for Cygnus A to estimate the pseudoscalar luminosity. The actual parameters may vary and hence our estimates may
only be qualitatively reliable.

We point out that we are not considering the standard Peccei-Quinn 
axion in this section. Rather, we look at a generic pseudoscalar  
whose mass and couplings to visible matter are unrelated to each other. The 
present bound on the axion mass is relatively large. For such a large mass, 
the conversion probability of axion into photon is found to be negligible. 
Instead, here we assume that $m_{\phi} \lesssim \omega_p$, where $\omega_p$ is
the typical plasma frequency in the radio lobes. In this case the conversion 
probability may be significant. 

The pseudoscalar-photon mixing phenomenon in background magnetic field has
been analyzed in great detail in the literature \cite{Sikivie83,Sikivie85,Maiani,RS88,Bradley,CG94,sudeep,Ganguly,Ganguly09,Csaki02,sroy,MirizziCMB,MirizziTEV,Song,Gnedin,Agarwal}. 
This phenomenon has also been used to impose stringent limits on
the pseudoscalar-photon coupling \cite{Rosenberg,Vysotsky,Dicus2,Dearborn86,Raffelt87,Raffelt88,Turner,Mohanty,Brockway,Grifols,CAST1,CAST2,jaeckel,Robilliard,zavattini,Rubbia}. 
A beam of photons, passing through background magnetic field, would convert 
partially into pseudoscalars due to this mixing phenomenon. 
The mixing probability increases with frequency. At very high frequencies 
the mixing probability is very large and hence the flux of pseudoscalars
produced may be comparable to the incident photon flux. As the pseudoscalar
flux becomes sizeable, we expect significant pseudoscalar to photon 
conversion. Eventually we expect a beam containing roughly equal number
of photons and pseudoscalars. This is true as long as the 
extinction of photons is negligible in the medium. However if extinction
is significant, we may obtain larger pseudoscalar luminosity, 
even if the incident beam consists entirely of photons.
The extinction coefficient for AGN atmospheres is not known. Here we shall
assume that the extinction is of same order of magnitude in comparison 
to what is observed in the host galaxies
in the case of high redshift supernovas \cite{Riess}. Here the visual 
extinction coefficient is extracted from the observed light curve for these
supernovas. In the present case, in order to compute the pseudoscalar flux 
at visible frequencies, we need to extrapolate the extinction coefficient 
to ultraviolet
frequencies. This is because a source at high redshift must emit UV light 
so that the radiation received by us is in the visible band. The extinction 
depends approximately linearly on the frequency of the  photons. Hence, we obtain the extinction at UV frequencies by suitably rescaling the extinction observed at the visible region in supernovae data \cite{Riess} at large redshifts.

We next briefly review the formalism for pseudoscalar-photon mixing in a 
uniform background, to the case where the medium causes extinction of photons. 
An earlier discussion of this phenomenon may be found in \cite{Csaki}. Using the notations used in \cite{sudeep}, we write the differential equation of mixing with extinction, ignoring the longitudinal component of the photons and the mixing of transverse components thereof \cite{sudeep}, as follows,
	\begin{equation}
	\left(\omega^2 + \partial^2_z\right)
	\left[\begin{array}{cc}
	A_{||}(z)\\
	\phi(z)
	\end{array}\right] = M
	\left[\begin{array}{cc}
	A_{||}(z)\\
	\phi(z)
	\end{array}\right]\,.
	\label{eq:mstart}
	\end{equation}
This equation describes the mixing of the parallel component of the electromagnetic field with the pseudoscalar $\phi$. The perpendicular component $A_{\perp}$ does not mix with $\phi$. The  ``mass matrix" or the ``mixing matrix" in Eq. (\ref{eq:mstart}) can be written as,

	\begin {equation}
	M = \left[
	\begin{array}{cc}
	\omega_p^2 + i\Gamma(\omega) & -g_{\phi}\mathcal{B_T}\omega \\
	-g_{\phi}\mathcal{B_T}\omega & m_{\phi}^2 \\
	\end{array}
	\right]
	\end{equation}
	
	\par\noindent
where $\mathcal{B_T}$ is the transverse component of the background magnetic 
field and $\Gamma(\omega)$ describes the attenuation of photons due to their 
extinction in the medium. We note that the extinction of light is modelled 
with a parameter called optical depth, $\tau_{\nu}$, such that the 
intensity $I_\nu(z) = I_\nu(0) e^{-\tau_\nu}$, where $z$ is the thickness
of the medium. In general the optical depth increases linearly with frequency.
Hence we may assume $\tau_\nu = K\omega$, where $K$ is a constant. 
In our formulation, the exponential decay parameter is $\frac{\Gamma z}{2\omega}$, at leading order. Equating this to $\tau_\nu$ we find 
$\Gamma = {2\omega^2 K \over z}$. 
As discussed above, we shall fix the value
of $K$ by assuming that at visual frequencies the extinction is similar to that observed for high redshift supernovas \cite{Riess}.
This leads to $\Gamma \approx 2.8\times 10^{-28}\omega^2(\tau/0.5)$, 
where $\omega$ is expressed in Hz and the visual extinction $A_V = (2.5\log_{10}e)\tau$. 

We can solve the equations, Eq. \ref{eq:mstart}, by diagonalizing 
the matrix $M$. The eigenvalues of this matrix, $\lambda_+$ and $\lambda_-$ 
may be expressed as, 
\begin{equation}
\lambda_\pm = {1\over 2}\left[\Omega_p^2 + m_\phi^2 \pm\sqrt{(\Omega_p^2 + m_\phi^2)^2 - 4(\Omega_p^2 m_\phi^2 - g_\phi^2 \mathcal{B}_T^2\omega^2) } \right]
\end{equation}
where $\Omega_p^2 = \omega_p^2+i\Gamma$.
We assume the boundary condition, $\phi(0) = 0$, and find the final result 
	\begin{eqnarray}
	\label{last}
	A_{||}(z) &=& {1\over ad-bc}\left[ad~e^{(iz\sqrt{\omega^2 - \lambda_+})} - bc~e^{(iz\sqrt{\omega^2 - \lambda_-})}\right] A_{||}(0) \nonumber \\
	\phi(z) &=& {bd\over ad-bc}\left[ e^{(iz\sqrt{\omega^2 - \lambda_+})} - e^{(iz\sqrt{\omega^2 - \lambda_-})}\right]A_{||}(0)\,,
	\end{eqnarray}
where $a = (\lambda_+-m_\phi^2)/\sqrt{N_+}$, $b = -g_\phi B_T\omega/\sqrt{N_+}$, $c = g_\phi B_T\omega/\sqrt{N_-}$, 
$d = (\Omega_p^2 - \lambda_-)/\sqrt{N_-}$. Here $N_+$ and $N_-$ are 
normalization factors which cancel out in the final expressions. 
The perpendicular component of the electromagnetic wave is given by,
\begin{equation}
A_\perp(z) = A_\perp(0)e^{iz\sqrt{\omega^2-\Omega_p^2}}
\label{eq:Aperp}
\end{equation}

Using Eq. \ref{last} and Eq. \ref{eq:Aperp} we can compute the photon and pseudoscalar flux emerging
out of the AGN atmosphere. In making this calculation we assume parameters 
corresponding to the Cygnus A radio lobe. Hence we set the plasma density 
$n_e = 10^{-4}$ cm$^{-3}$ and magnetic field $B_T = 4\times 10^{-4}$ G.
 We assume the pseudoscalar photon coupling, 
$g_{\phi\gamma\gamma} = 10^{-10}$ GeV$^{-1}$ and the pseudoscalar mass is
set to zero.  

\begin{figure}[t]
	\centering
	\includegraphics[]{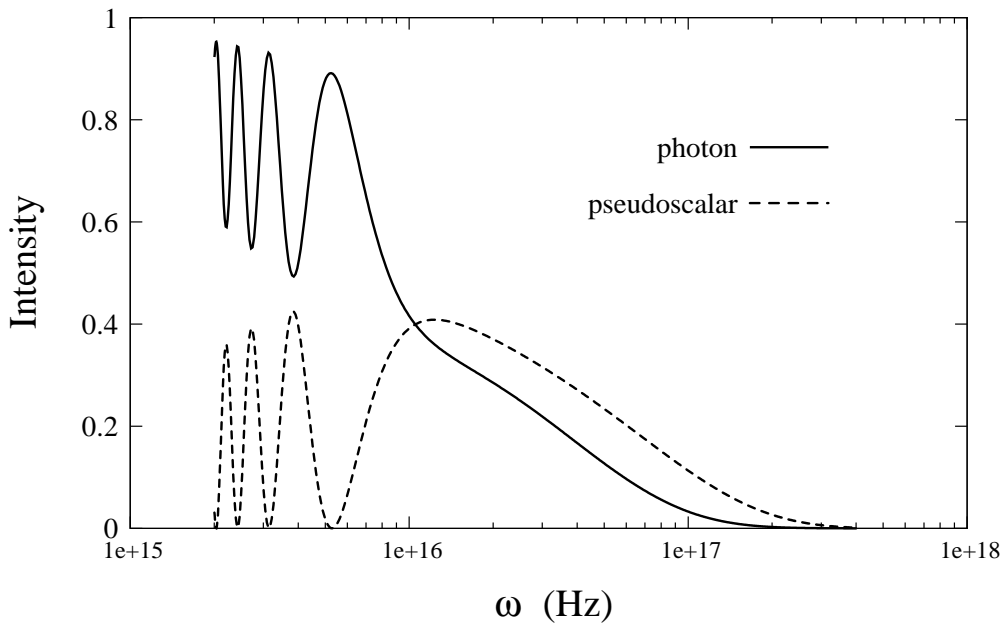}
	\caption{The pseudoscalar and photon intensity as a function of
	the frequency.}
	\label{fig:omega}
	\end{figure}

\begin{figure}[t]
	\centering
	\includegraphics[]{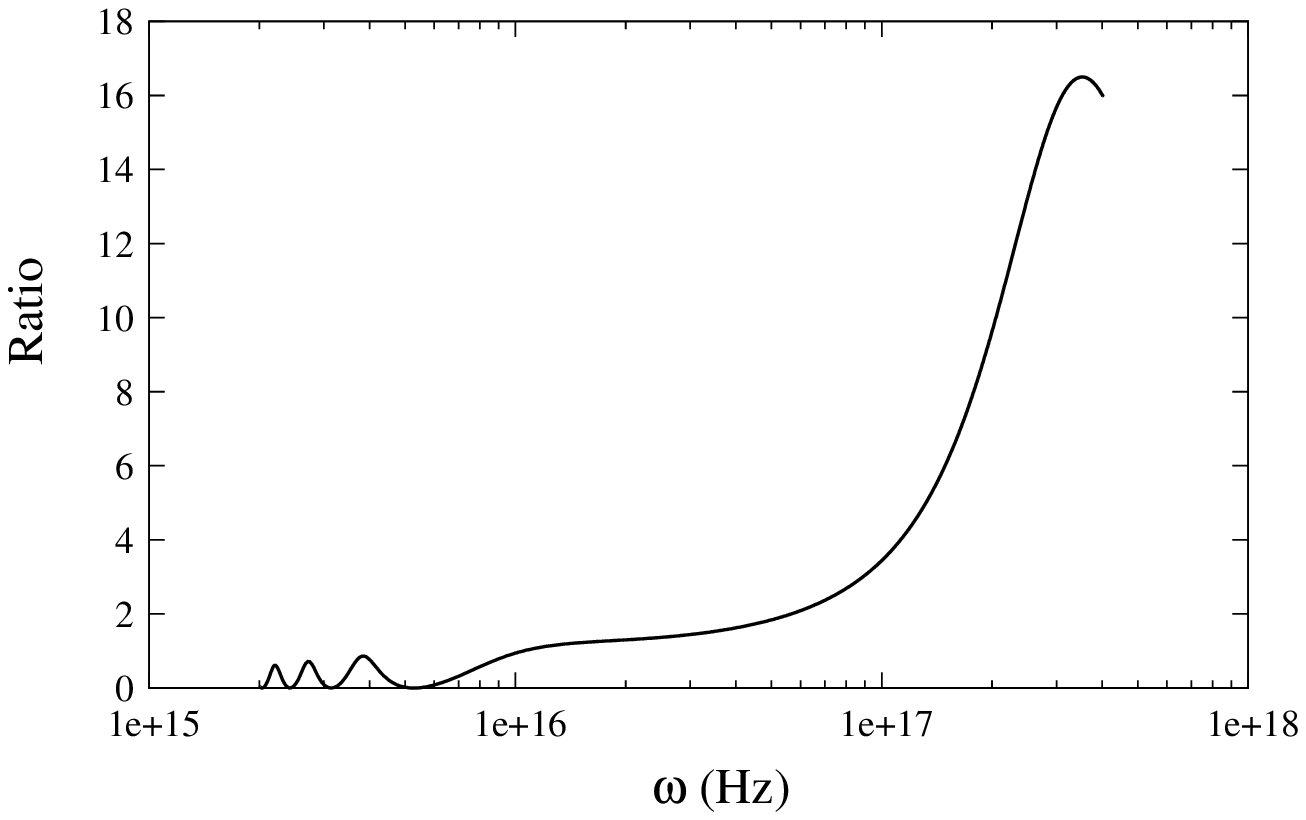}
	\caption{The ratio of pseudoscalar to photon intensity as a 
	function of the frequency.}
	\label{fig:ratio}
	\end{figure}

In Fig. \ref{fig:omega} we show the pseudoscalar and photon intensity as a
function of the frequency setting the distance equal to 10 Kpc. 
Here we have set the extinction parameter $\tau=0.1$ and $\omega$ represents
the frequency at source. In this plot we have normalized the intensity
such that the photon intensity is unity before entering the AGN atmosphere. 
In Fig. \ref{fig:ratio} we show the ratio of the pseudoscalar to photon 
intensity as a function of frequency.
We find that the pseudoscalar intensity is significantly larger in comparison
to the photon intensity at higher frequencies. For the parameters 
chosen, the pseudoscalar intensity is a factor of two or three larger 
than the photon intensity for $\omega=5\times 10^{16}$ to $10^{17}$ Hz. 
For larger frequencies the pseudoscalar flux
may be an order of magnitude higher in comparison to the photon flux. 
However here the overall flux may be very small. 

In our calculations we have set the pseudoscalar mass to zero. If the
pseudoscalar mass is comparable to the plasma density of the medium then
there is also the possibility of resonant mixing of pseudoscalars with 
photons \cite{sudeep}. In this case the photon to pseudoscalar conversion
is considerably enhanced and hence the pseudoscalar flux from AGNs
may be significantly higher.

\section{Alignment of Quasar Polarizations}
We next briefly address the issue of alignment of optical polarization
from quasars in the direction of the Virgo supercluster. This region is
labelled as A1 in \cite{huts98,huts01}. Here we limit ourselves to a qualitative
explanation. A detailed quantitative analysis is postponed for future
research.

The alignment effect may in principle be explained
by the conversion of photons into pseudoscalars in the Virgo supercluster
magnetic field \cite{huts98,huts01,JPS02}. However, as mentioned in the introduction, 
this does not consistently explain the data due to the observed 
distribution of polarization of the RQ and O quasars \cite{huts98,huts01}. 
The polarization
distribution of these quasars is found to peak at very low values. 
In contrast the distribution of BAL quasars is observed to be much broader.
The difference between these two classes of quasars is seen in all directions
including the A1 region.
The magnitude of the systematic effect required to explain alignment
is sufficiently large
that it would completely distort the distribution of RQ and O quasars. 
We may alternatively
consider the possibility that quasars emit a significant amount of 
pseudoscalar flux, as found in the previous section. The pseudoscalar flux is
assumed to be larger than the photon flux. In this case the 
alignment may be explained in terms of the conversion of pseudoscalars 
into photons in the Virgo supercluster. This will lead to a linear 
polarization aligned along the transverse component of the background magnetic
field. Furthermore we assume that the
ratio of the pseudoscalar to photon flux is much larger for the BAL quasars
in comparison to the RQ and O quasars. Hence the systematic polarization
effect due to pseudoscalar-photon mixing would be smaller for RQ and O
quasars in comparison to BAL quasars. Since the RQ and O quasars in general 
have smaller degree of polarization, this may be sufficient to explain
their alignment. In contrast the BAL quasars would get a larger contribution
due to pseudoscalar-photon mixing, which is required due to their larger 
intrinsic polarization. Hence we find that the observed alignment may be
consistently explained if the quasars emit pseudoscalars.

\section{Conclusions}
	
\par\noindent In this paper, we have found that the luminosity of pseudoscalars from the AGN accretion disk due to Compton, Bremsstrahlung and  Primakoff channels is very small in comparison to the photon luminosity. However, the photons in visible and ultraviolet frequencies may convert to pseudoscalars outside the accretion disk due to pseudoscalar-photon mixing in the background magnetic field. Taking extinction of photons into account and using the current limit on the pseudoscalar-photon coupling, we find that the pseudoscalar flux produced by this process is relatively large. For ultraviolet frequencies, which would be observed in the visible range on earth, this flux may dominate the  photon flux. A large pseudoscalar flux may provide a consistent explanation for the large scale coherent orientation of the visible polarizations from quasars.
	\medskip
	\\

	\section{Acknowledgement}
We thank Suman Bhattacharya for collaboration during 
the initial stages of this work.


\begin{thebibliography}{unsrt}

	\bibitem{bmp} B.M. Peterson,~\textit{An Introduction To Active Galactic Nuclei}, \textsf{Cambridge Univ. Press}, (1997).
	\bibitem{kori} K. T. Korista et. al.,~\textit{Astrophys. J. Suppl.},~\textbf{97}, \textsf{285}, (1995).
\bibitem{PQ1} R. D. Peccei and H. Quinn, {\it Phys. Rev. Lett.} {\bf
38}, 1440 (1977). 
\bibitem{PQ2} R. D. Peccei and H. Quinn, {\it Phys. Rev.} {\bf D16}, 1791 (1977).
\bibitem{Weinberg} S. Weinberg, {\it Phys. Rev. Lett.}  {\bf 40},
223 (1978). 
\bibitem{Wilczek} F. Wilczek, {\it Phys. Rev. Lett.},  {\bf 40},
279 (1978).
\bibitem{McKay} D. McKay, Phys. Rev. {\bf D16}, 2861 (1977).
\bibitem{Kim1} J. E. Kim, Phys. Rev. Lett. {\bf 43}, 103 (1979).
\bibitem{Dine} M. Dine, W. Fischler and M. Srednicki,
{\it Phys. Lett.},  {\bf B104}, 199 (1981).
\bibitem{Zhitnitsky} A. R. Zhitnitsky, \textit{Sov. Jour. Nucl. Phys.}, \textbf{31}, \textsf{260}, (1980).
\bibitem{McKay2} D. McKay and H. Munczek, Phys. Rev. {\bf D19}, 985 (1979).
\bibitem{Kim} J. E. Kim, Phys. Rept. {\bf 150}, 1 (1987).
	\bibitem{huts98} D. Hutsem\'{e}kers, \textit{Astron. Astrophys.},~\textbf{332}, \textsf{410}, (1998).
	\bibitem{huts01} D. Hutsem\'{e}kers and H. Lamy, \textit{Astronomy and Astrophysics}, \textbf{367}, \textsf{381}, (2001).
	\bibitem{huts02} D. Hutsem\'{e}kers, R. Cabanac, H. Lamy and D. Sluse, 
\textit{Astronomy and Astrophysics}, \textbf{441}, \textsf{915}, (2005).
	\bibitem{JNS04} P. Jain, G. Narain and S. Sarala, \textit{MNRAS}, \textbf{347}, \textsf{394}, (2004).
	\bibitem{JP04} J. P. Ralston and P. Jain, Int. J. Mod. Phys. 
	\textbf{D13}, 1857 (2004).   
	\bibitem{JPS02} P. Jain, S. Panda and S. Sarala, \textit{Phys. Rev.} 
	\textbf{D66},
	\textsf{085007}, (2002).
	\bibitem{kolb} E. Kolb and M. S. Turner, \textit{The Early Universe}-\textsl{Chapter 10}, \textbf{Westview Press}, \textsf{2nd Ed.}, (1994).
	\bibitem{kmw} L. M. Krauss et. al.,~\textit{Phys. Lett.} \textbf{B144}, \textsf{391}, (1984).
\bibitem{Dicus1} D. Dicus, E. Kolb, V. Teplitz and R. Wagoner,
Phys. Rev.  {\bf D 18}, 1829 (1978).
	\bibitem{fukugita} M. Fukugita, S. Watamura and M. Yoshimura, 
\textit{Phys. Rev.} \textbf{D26}, \textsf{1840}, (1982).
	\bibitem{PDG} C. Amsler et al. (Particle Data Group), Physics Letters 
	\textbf{B667}, 1 (2008) 
	\bibitem{ggrflt} G. G. Raffelt, \textit{Lect. Notes Phys.}, \textbf{741}, \textsf{51}, (2008). 
	\bibitem{king} ~J. Frank, A. King \& D. Raine, ~\textit{Accretion Power In Astrophysics},~\textsf{Cambridge University Press}, (2002).
\bibitem{Sikivie83} P. Sikivie, {\it Phys. Rev. Lett.}  {\bf 51}, 1415 (1983).
\bibitem{Sikivie85} P. Sikivie, 
{\it Phys. Rev.} {\bf D32}, 2988 (1985).
  \bibitem{Maiani} L. Maiani, R. Petronzio and E. Zavattini,
  {\it Phys. Lett.}   {\bf B175}, 359 (1986).
  \bibitem{RS88} G. G. Raffelt and L. Stodolsky, {\it Phys. Rev.}  {\bf D37},
  1237 (1988).
  \bibitem{Bradley} R. Bradley {\it et al}, Rev. Mod. Phys.  {\bf 75},
  777 (2003).
  \bibitem{CG94} E. D. Carlson and W. D. Garretson, {\it Phys. Lett.}
   {\bf B336}, 431 (1994).
   \bibitem{sudeep} S. Das, P. Jain, J. P. Ralston and R. Saha, JCAP {\bf 0506},
   002 (2005).
   \bibitem{Ganguly} A. K. Ganguly, Annals of Physics, {\bf 321}, 1457 (2006).
   \bibitem{Ganguly09} A. K. Ganguly, P. Jain and S. Mandal, Phys. Rev. 
   {\bf D79}, 115014 (2009).
\bibitem{Csaki02} C. Csaki, N. Kaloper, J. Terning, {\it Phys. Rev. Lett.}
{\bf 88}, 161302, (2002);  Phys. Lett. {\bf B 535}, 33, (2002).
\bibitem{sroy} Y. Grossman, S. Roy and J. Zupan, Phys. Lett. {\bf B 543}, 23 (2002).
\bibitem{MirizziCMB} A. Mirizzi, G. G. Raffelt and P. D. Serpico,
Phys. Rev. {\bf D 72}, 023501 (2005).
\bibitem{MirizziTEV} A. Mirizzi, G. G. Raffelt and P. D. Serpico,
Phys. Rev. {\bf D 76}, 023001 (2007).
\bibitem{Song} Y.-S. Song and W. Hu, Phys. Rev. {\bf D 73}, 023003 (2006).
\bibitem{Gnedin} Y. N. Gnedin, M. Yu. Piotrovich and T.M. Natsvlishvili, MNRAS,
{\bf 374}, 276 (2007).
\bibitem{Agarwal} N. Agarwal, P. Jain, D. W. McKay and J. P. Ralston,
Phys. Rev. D {\bf 78}, 085028 (2008).


\bibitem{Rosenberg} L. J. Rosenberg and K. A. van Bibber, {\it Phys.
Rep.}  {\bf 325}, 1 (2000).
\bibitem{Vysotsky} M. I. Vysotsky, Ya. B. Zeldovich, M. Yu. Khlopov and V. M. 
Chechetkin, JETP Lett. {\bf 27}, 502 (1978).  
\bibitem{Dicus2} D. Dicus, E. Kolb, V. Teplitz and R. Wagoner,
 Phys. Rev.  {\bf D 22}, 839 (1980).
\bibitem{Dearborn86}
D. Dearborn, D. Schramm and G. Steigman, Phys. Rev. Lett.  {\bf 56}, 26
(1986). 
\bibitem{Raffelt87} 
G. G. Raffelt and D. Dearborn, Phys. Rev.  {\bf D 36}, 2211 (1987).
\bibitem{Raffelt88} 
G. G. Raffelt and D. Seckel, Phys. Rev. Lett.  {\bf 60}, 1793 (1988);
\bibitem{Turner} 
M. Turner, Phys. Rev. Lett.  {\bf 60}, 1797 (1988).
\bibitem{Mohanty} S. Mohanty and S. N. Nayak, {\it Phys. Rev. Lett.}
{\bf 70}, 4038, (1993).
\bibitem{Brockway} J. W. Brockway, E. D. Carlson and G.G. Raffelt,
{\it Phys. Lett.}  {\bf B 383}, 439 (1996).
\bibitem{Grifols} J. A. Grifols, E. Masso and R. Toldra, Phys. Rev. Lett.  {\bf 77}, 2372 (1996).
\bibitem{CAST1} K. Zioutas et al., Phys. Rev. Lett.  {\bf 94}, 121301 (2006).
\bibitem{CAST2}
S. Andriamonje et al., JCAP  {\bf 0704}, 010 (2007).
\bibitem{jaeckel} J. Jaeckel, E. Masso, J. Redondo, A. Ringwald and F. Takshashi, Phys. Rev. {\bf D 75}, 013004 (2007).
\bibitem{Robilliard} C. Robilliard, R. Battesti, M. Fouche, J. Mauchain,
A.-M. Sautivet, F. Amiranoff and Carlo Rizzo,
Phys. Rev. Lett.  {\bf 99}, 190403 (2007).
\bibitem{zavattini} E. Zavattini {\it et al.} Phys. Rev. {\bf D 77}, 032006 (2008).
\bibitem{Rubbia} A. Rubbia and A. S. Sakharov, Astropart. Phys. {\bf 29},
20 (2008).



\bibitem{Riess} A. G. Riess {\it et al}, Astrophysical Journal,	{\bf 607}, 
665 (2004).
\bibitem{Csaki} C. Csaki, N. Kaloper, M. Peloso and J. Terning, JCAP {\bf 0305},
005 (2003).
	\end{thebibliography}
	\end{document}